\documentclass[letterpaper,twocolumn]{jpsj3}
\usepackage{txfonts}

\title{Electrically-excited Motion of Topological Defects in Multiferroic Materials}

\author{Maxim Mostovoy\thanks{m.mostovoy@rug.nl}}
\inst{Zernike Institute for Advanced Materials, University of Groningen,\\ Nijenborgh 4, 9747 AG, Groningen, The Netherlands } 

\abst{Topological magnetic defects in multiferroic materials acquire an electric charge or dipole moment due to the inverse Dzyaloshinskii-Moriya mechanism. This magnetoelectric coupling makes possible to excite large-amplitude collective motion of topological magnetic textures with an oscillating electric field. Here, I discuss  electric excitation of a polar optical mode in a vortex-antivortex crystal and the electrically-induced spin precession in a magnetic skyrmion, which gives rise to rotation of skyrmions around each other and a translational motion of skyrmion-antiskyrmion pairs. The electric manipulation of magnetic topological defects in Mott insulators can find applications in magnetoelectric  memory and logical devices.
}

\begin{document}
\maketitle

\section{Introduction}
\label{sec:introduction}
Non-collinear magnetism gives rise to many interesting phenomena, such as the topological Hall effect, \cite{Bruno2004,Neubauer2009,Lee2009} multiferroicity,  \cite{Kimura2007,Cheong2007,Khomskii2009,Tokura2010} and electromagnons.\cite{Pimenov2006,Katsura2007,Aguilar2009} Igor Dzyaloshinskii understood the relation between non-collinear magnetism and broken inversion symmetry of the crystal lattice.  \cite{Dzyaloshinskii} He noted that non-centrosymmetric magnets allow for the so-called Lifshitz invariants in their free energy, $M_i \overleftrightarrow{\partial_j} M_k = M_i \partial_j M_k - M_k \partial_j M_i$, where $\mathbf{M}$ is  the magnetization vector  and the spatial indices $i,j,k$ depend on crystal symmetry. Toru Moriya explained the microscopic mechanism behind these invariants: the spin-orbit coupling gives rise to electron hopping with spin-flip, which leads to effective interactions proportional to the vector product of spins, favoring non-collinear magnetic states.  \cite{Moriya}  These Dzyaloshinskii-Moriya interactions (DMI) involve relativistic effects that are relatively weak in compounds with magnetic $3d$ transition metal ions. They explain the origin of long-wavelength spirals found in non-centrosymmetric magnets. 

Bogdanov and co-workers showed that DMI  can stabilize more complex spin textures with non-trivial topology, such as skyrmions and antiskyrmions. \cite{Bogdanov1989,Bogdanov2002} The observation of skyrmion crystal in MnSi with a non-centrosymmetric cubic lattce\cite{Muelbauer2009,Yu2010} opened a new research field - skyrmionics, for studying skyrmions in bulk and multilayer materials, in particular, their dynamics resulting from  interactions with electrons and magnons, with the aim to use topological magnetic defects for storage and processing of information.\cite{Nagaosa2013,Fert2013,Back2020}

Non-collinear spiral states can also be found in centrosymmetric magnets, where they emerge as a compromise between competing Heisenberg exchange interactions.   \cite{Villain1959,Yoshimori1959} Recent theoretical studies showed that frustrated magnets can host skyrmion crystals and other multiply-periodic states, as well as metastable isolated skyrmions and merons.  \cite{Okubo2012,Leonov2015,Hayami2016,Kharkov2017} Non-collinear magnetism can also originate from the spin-density-wave instability of the electron Fermi surface and long-ranged interactions mediated by itinerant electrons in magnetic conductors. \cite{Hayami2014,Motome2017,Mostovoy2005,Pekker2005,Azhar2017} Crystals of nanosized skyrmions resulting in  giant Topological Hall and Nernst effects have recently been observed in a number of intermetallic compounds with hexagonal and tetragonal crystal lattices.\cite{Kurumaji2019,Hirschberger2019,Khan2020} The itinerant magnet, SrFeO$_3$, with a centrosymmetric cubic perovskite lattice, shows magnetic states with several coexisting spin spirals, one of which was identified with a three-dimensional crystal of magnetic hedgehogs and antihedgehogs.\cite{Ishiwata2020} In contrast to chiral magnets, the direction of spin rotation in spiral states of centrosymmetric magnets is arbitrary, which allows for more versatile spin textures and new  collective degrees of freedom of topological magnetic defects. 

The choice of spin-rotation direction at the transition into a spin-spiral state  spontaneously breaks inversion symmetry of a centrosymmetric magnet, allowing for an electric polarization in the magnetically ordered state. \cite{Baryachtar1983,Katsura2005,Sergienko2006,Mostovoy2006} Recent studies of multiferroic materials led to discovery of a large number of Mott insulators with competing exchange interactions and electric polarization induced by a spin-spiral ordering. \cite{Kimura2007,Cheong2007,Khomskii2009,Tokura2010} This phenomenon is often referred to as the inverse Dzyaloshinskii-Moriya effect. Microscopically, the magnetically-induced polarization originates from the sensitivity of DMI to geometry of metal-ligand-metal bonds and, in particular, to polar shifts of ions and a redistribution of the electron density induced by a non-collinear spin ordering. More generally, electric polarization can be induced by linear electric-field dependence of all parameters in the spin Hamiltonian, e.g., Heisenberg exchange constants, \cite{Chapon2006,SergienkoSen2006,Bulaevskii2008} magnetic anisotropy \cite{Murakawa2012}  and $g$-tensors.  \cite{Scaramucci2012}  The coupling between spin and charge degrees of freedom in Mott insulators allows for the magnetic control of electric polarization and leads to many unconventional phenomena, such as the giant magnetocapacitance \cite{Goto2004} and excitation of magnons by the electric component of light resulting in non-reciprocal optical phenomena.   \cite{Kezsmarki2011,Takahashi12} The magnetoelectric coupling induces an electric polarization at magnetic domain walls\cite{Baryachtar1983} and an electric charge in the core of a magnetic vortex. \cite{Mostovoy2006}

This paper written in memory of I. Dzyaloshinskii and T. Moriya focuses on the interplay between topological and magnetoelectric properties of spin textures and the resulting electrically-driven  dynamics of topological magnetic defects, such as skyrmions and merons, in multiferroic Mott insulators. In Sect.~\ref{sec:polarization} I discuss phenomenological description of the magnetoelectric coupling originating from the inverse Dzyaloshinskii-Moriya mechanism and the relation between electric and topological charge densities of inhomogeneous spin textures. An interesting consequence of this relation is the `optical phonon mode' excited by an oscillating electric field in a vortex-antivortex crystal (see Sect.~\ref{sec:plasmon}). In Sect.~\ref{sec:spinrotation}, I discuss the vorticity and helicity dependence of the skyrmion electric dipole moment, which makes possible to rotate spins with an oscillating electric field. The coupling between the skyrmion helicity and center-of-mass dynamics leads to rotation of two skyrmions around each other (Sect.~\ref{sec:skskpair}) and to a translational motion of skyrmion-antiskyrmion pairs in an applied electric field (Sect.~\ref{sec:skaskpair}). Section~\ref{sec:conclusions} contains discussion and summary. 

\section{Electric polarization and charge induced by a magnetic texture}
\label{sec:polarization}

The form of magnetoelectric coupling describing the electric polarization induced by an inhomogeneous magnetic texture depends on crystal symmetry. Consider a centrosymmetric magnet with a three-fold, four-fold or six-fold symmetry axis ($z$-axis), which allows for the energy term linear in the electric field $\mathbf{E}$,
\begin{align}
E_{\rm me} & =-\lambda_{\|} E_i \left(m_j \partial_j m_i - m_i \partial_j m_j\right) - 
\lambda_{\perp} E_z \left(m_j \partial_j m_z - m_z \partial_j m_j
\right)
\nonumber\\
& = -\lambda_{\|} E_i m_j \overleftrightarrow{\partial_j} m_i 
-\lambda_{\perp} E_z m_j \overleftrightarrow{\partial_j} m_z,
\label{eq:fmexy}
\end{align}
where the unit vector $\mathbf{m}(x,y)$ describes the magnetization direction varying slowly in the $xy$-plane at the lattice constant scale, $i,j$ denote in-plane  ($x$ or $y$) directions with implied summation over repeated indices, and $\lambda_{\|}/\lambda_\perp$ is the coupling constant for the electric field parallel/perpendicular to the $xy$-plane. This magnetoelectric coupling has the form of Lifshitz invariant multiplied by an electric field component. It involves two  pairs of polar and axial vectors with the same indices and, hence,  is invariant under inversion, all vertical mirrors and rotations around the $z$-axis through an arbitrary angle. This high symmetry makes the magnetoelectric coupling Eq.(\ref{eq:fmexy}) compatible with symmetries of trigonal, tetragonal and hexagonal magnets. Other magnetoelectric coupling terms that might be allowed by crystal symmetry are not considered here. 

The in-plane electric polarization is given by
\begin{equation}
P_i = - \frac{\partial E_{\rm me}}{\partial E_i} = \lambda_{\parallel} m_j \overleftrightarrow{\partial_j} m_i.
\end{equation}
For a spiral with a wave vector $\mathbf{Q}$ parallel to the $xy$-plane and spins rotating around the $z$-axis, $\mathbf{m} = (\cos \phi, \sin \phi,0)$, where $\phi =\mathbf{Q} \cdot \mathbf{x} + \phi_0$, the polarization is  orthogonal to $\mathbf{Q}$: $\mathbf{P} = \lambda_{\|} 
\left[
\hat{\mathbf{z}} \times \mathbf{Q} 
\right]$. Similarly, a spiral with spins rotating in a vertical plane induces an out-of-plane electric polarization, in accordance  with the inverse Dzyaloshinskii-Moriya or spin current mechanism,  \cite{Katsura2005,Sergienko2006,Mostovoy2006} which  Eq.(\ref{eq:fmexy}) describes phenomenologically.

The density of electric charge induced by $\mathbf{m}(x,y)$ is 
\begin{equation}
\rho_{\rm e} = - \partial_i P_i = 2 \lambda_{\|} \sin \theta \cos \theta(\partial_x\theta\partial_y\phi -\partial_y\theta\partial_x\phi).
\end{equation}
Comparing the last equation with the topological (skyrmion) charge density  (see e.g. 
Ref.~\citen{Nagaosa2013}),
\begin{equation}
\rho_{\rm sk} = \frac{1}{4\pi}(\mathbf{m} \cdot \partial_x 
\mathbf{m} \times \partial_y \mathbf{m} ) =  \frac{1}{4\pi} \sin \theta (\partial_x \theta \partial_y \phi - \partial_y\theta\partial_x\phi),
\end{equation}
we find
\begin{equation}
\rho_{\rm e} = 8 \pi \lambda_{\|} \rho_{\rm sk} m_z.
\end{equation}
Note that the skyrmion charge density, which plays the role of  $z$-component of an effective magnetic field,\cite{Bruno2004,Neubauer2009,Lee2009} is odd under time reversal, whereas the charge density is even.

Next we consider a rotationally symmetric spin texture, e.g. a vortex or skyrmion, with $\theta = \theta(r)$ and $\phi = {\rm v} \varphi + \chi$, where $(r,\varphi)$ are the polar coordinates in the $xy$-plane, integer ${\rm v}$ is the vorticity (winding number) and $\chi$ is the helicity angle. \cite{Nagaosa2013} The total skyrmion and electric charges of such a texture are given by
\begin{equation}
Q_{\rm sk} = \frac{{\rm v}}{2}\left(m_z(0)-m_z(\infty)\right),\quad
Q_{\rm e} = 2 \pi \lambda_{\|} {\rm v}\left(m_z^2(0)-m_z^2(\infty)\right).
\label{eq:charges}
\end{equation}
A skyrmion with $m_z(0) = -m_z(\infty) = \pm 1$ and $Q_{\rm sk} = {\rm v} m_z(0)$ has zero electric charge. A meron with $m_z(0) = \pm 1$, $m_z(\infty) = 0$ has topological charge $Q_{\rm sk} = \frac{{\rm v}}{2}m_z(0)$ and electric charge 
\begin{equation}
Q_{\rm e} = 2 \pi \lambda_{\|} {\rm v}.
\end{equation}
A meron is a vortex with the vorticity ${\rm v}$ and a soft core: the singularity at $r = 0$ is avoided by an out-of-plane orientation of $\mathbf{m}$. Outside the core, magnetization lies in the $xy$-plane and  $\mathbf{P} = \lambda_{\|} \left[\hat{\mathbf{z}} \times \mathbf{\nabla} \phi \right]$. The total electric charge can then be expressed as an integral over a contour $\Gamma$ around the vortex core:
\begin{equation}
Q_{\rm e} = - \oint_\Gamma d\mathbf{S} \cdot \mathbf{P} = \lambda_{\|}  \oint_\Gamma d\mathbf{x} \cdot \mathbf{\nabla} \phi = 2 \pi \lambda_{\|} {\rm v},
\end{equation}
where $d\mathbf{S}$ is the surface element normal to the contour and $d \mathbf{x}$ is the displacement along the contour. The electric charge is contour-independent, since it is confined to the vortex core. Charge conservation implies that the charge $- Q_{\rm e}$ is pushed to infinity (sample edges). Compact topological defects with a finite energy, such as skyrmions, cannot pump the electric charge to infinity and, hence, have zero total electric charge. 

\section{Electromagnon in vortex-antivortex crystal}
\label{sec:plasmon}

\begin{figure}
	\centering
	\includegraphics[width=0.9\linewidth]{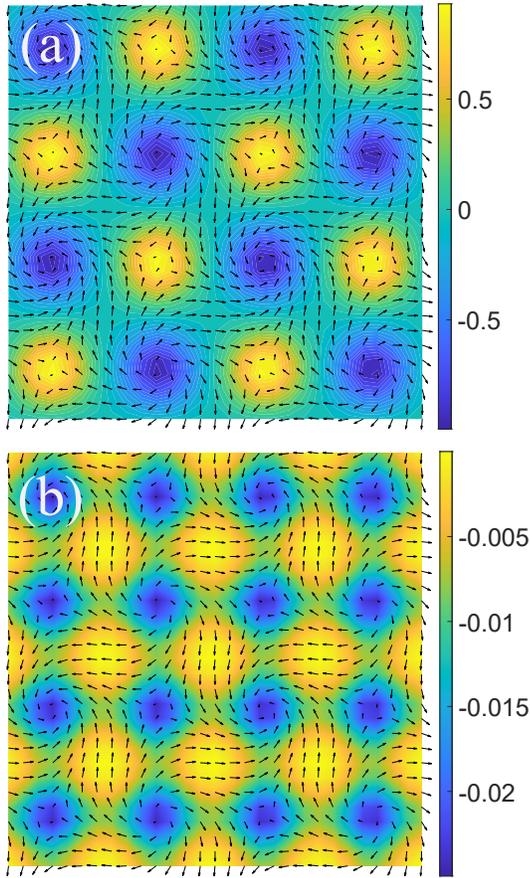}
	\caption{(Color online) (a) The vortex-antivortex crystal on a square spin lattice with competing Heisenberg exchange interactions. The in-plane spin-components are indicated by arrows; the out-of-plane spin component is shown by contour plot. (b) The same crystal as in panel (a), but with the contour plot showing the topological charge density. Both vortices and antivortices are merons with topological charge $Q_{\rm sk} = -1/2$.}
	\label{fig:Meroncrystal}
\end{figure}

Electrically charged cores make possible to move magnetic vortices with an electric field. Consider a vortex-antivortex crystal  (see Fig.~\ref{fig:Meroncrystal}(a)) in a frustrated magnet on a square lattice  with the energy
\begin{align}
	E=
	\sum_{\langle n,m\rangle} J_{nm}(\mathbf{m}_n\cdot \mathbf{m}_m)-
	\frac{K}{2}\sum_n(m_{nz})^2.  
	\label{eq:esquare}
\end{align}
The three exchange constants in the first term describe the ferromagnetic nearest-neighbor interaction, $J_1 = -1$, antiferromagnetic next-nearest-neighbor interaction, $J_2 = 0.3$, and the weak third-nearest-neighbor interaction, $J_3 = 0.0975$; the second term with $K = -0.05$ is an easy-plane anisotropy. The ground state state of the model for these parameters is ferromagnetic and the vortex-antivortex array is a metastable state stabilized in the vicinity of Lifshitz point\cite{Kharkov2017} by the competition between frustrated exchange interactions, favoring modulated magnetic states, and magnetic anisotropy, favoring a uniform ferromagnetic state. 

In the vortex core, $m_z < 0$, whereas in the antivortex core, $m_z > 0$ (see  Fig.~\ref{fig:Meroncrystal}(a)), so that according to Eq.(\ref{eq:charges}) both vortices and antivortices are merons with topological charge $Q_{\rm sk}=-1/2$. The distribution of  topological charge density is shown in  Fig.~\ref{fig:Meroncrystal}(b).  

\begin{figure}
	\centering
	\includegraphics[width=0.9\linewidth]{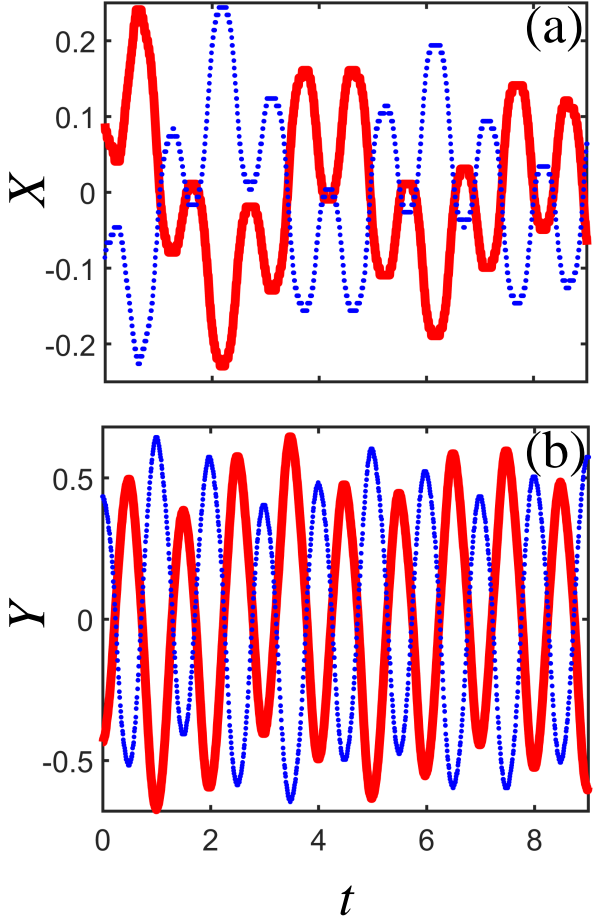}
	\caption{(Color online) (a) Time dependence of the displacements along the (a) $x$- and (b) $y$-direction of a vortex (solid line) and an antivortex (dotted line) in the vortex-antivortex crystal under an oscillating electric field applied along the $x$-axis. The $X$ and $Y$ coordinates of merons are measured in units of the lattice constant and the time $t$ is measured in units of the electric field oscillation period $T = \frac{2\pi}{\omega}$.}
	\label{fig:oscillations}
\end{figure}

We now add the coupling to electric field applied along the $x$ direction (see Eq.(\ref{eq:fmexy})): 
\begin{equation}
E_{\rm me}=  \lambda_{\|} E_x \sum_n \left[\mathbf{m}_n\times \mathbf{m}_{n+y}\right]_z,
\end{equation}
$(n,n+y)$ being a pair of neighboring sites along the $y$-direction. 
In the periodically varying electric field $E_x(t) = E_0 \cos (\omega t)$, the vortices and antivortices begin to oscillate. The vortex/antivortex center is defined as an interpolated position of a point where $m_z = -1/+1$. Figure~\ref{fig:oscillations} shows the time dependence of the displacements of vortices (solid line) and antivortices (dotted line) obtained by numerical solution of Landau-Lifshitz-Gilbert (LLG) equation with the Gilbert damping parameter $\alpha = 0.05$. Interestingly, the amplitude of shifts in the $y$-direction (panel (b)) is about three times larger than that along the $x$-direction (panel (a)). In addition, vortices and antivortices shift in opposite directions.

This peculiar dynamics can be understood from the fact that vortices and antivortices have opposite electric charges (see Eq.(\ref{eq:charges})) and, therefore, are pulled by the electric field in opposite directions. Furthermore, the gyrotropic dynamics of (anti)vortices implies that they move in the direction nearly perpendicular to the force. Thus, the electromagnon mode excited by the electric field is, essentially, a polar phonon mode in an ionic crystal, except that the electrically charged topological defects move in a strange direction.   

\section{Electrically induced spin rotation}
\label{sec:spinrotation}

In absence of an in-plane magnetic anisotropy, the skyrmion energy is independent of its helicity angle $\chi$, i.e. helicity is a zero mode. An electric field applied along the $z$-axis lifts this degeneracy making it possible to excite helicity dynamics electrically.

Due to the magnetoelectric coupling Eq.(\ref{eq:fmexy}), the skyrmion with a vorticity ${\rm v}$ and helicity angle $\chi$, described by
\begin{align}
\mathbf{m} & = (m_x,m_y,m_z)\nonumber\\
& =(\sin \theta(r)\cos({\rm v} \varphi + \chi),\sin \theta(r)\sin({\rm v} \varphi + \chi),\cos \theta(r)),
\label{eq:skyrmion}
\end{align}
induces the electric polarization along the $z$ axis
\begin{equation}
P_z 
= -\lambda_{\perp}\left[\frac{d\theta}{dr}+{\rm v}\frac{\sin 2\theta}{2r}\right] \cos \left(({\rm v}-1)\varphi + \chi\right).
\end{equation}
The last equation implies that (i) the net electric dipole moment $d_z = \int\!\!d^2x P_z$ is only nonzero for ${\rm v} = +1$ and (ii) $d_z$ depends on the helicity angle:
\begin{equation}
d_z(\chi) = D \cos\chi.
\label{eq:dz}
\end{equation} 
In particular, antiskyrmion (${\rm v} = -1$) and Bloch skyrmion ($ {\rm v} = +1, \chi = \pm \frac{\pi}{2}$) induce no elecric dipole moment, whereas N\'eel skyrmion with ${\rm v} = +1$ and  $\chi=0$ ($\chi=\pi$) has the electric dipole moment $+D$ ($-D$).

An interesting consequence of the helicity-dependence of $d_z$ is the rotation of spins around the $z$-axis induced by an oscillating electric field $E_z(t) = E_0 \cos(\omega t +\varphi_E)$. Consider a model of a frustrated triangular magnet \cite{Leonov2015} with the energy,
\begin{align}
	E=
	\sum_{\langle n,m\rangle}J_{nm}(\mathbf{m}_n\cdot \mathbf{m}_m)
	- \sum_n \left[h_z m_{nz}+\frac{K}{2}(m_{nz})^2 + E_{\rm me}
	\right],
	\label{eq:etriang}
\end{align}
where the first term describes competing Heisenberg exchange interactions: the nearest-neighbor ferromagnetic interaction $J_1 < 0$   and next-nearest-neighbor antiferromagnetic  interaction $J_2 > 0$; the second term is the Zeeman energy, and the third term is an easy-axis magnetic anisotropy with $K > 0$. The last term is a discrete version of the magnetoelectric coupling Eq.(\ref{eq:fmexy}):
\begin{equation}
E_{\rm me} = \frac{\lambda_\perp}{\sqrt{3}} E_z \sum_{n,\alpha} m_{nz}\left((\mathbf{m}_{n+e_\alpha}-\mathbf{m}_{n-e_\alpha})\cdot\mathbf{e}_{\alpha}\right).
\end{equation}
The unit vectors $\pm \mathbf{e}_{\alpha }$ ($\alpha = 1,2,3$) connect nearest-neighbor sites of a triangular lattice with the lattice constant $a = 1$: $\mathbf{e}_{1}=\hat{\mathbf {x}}$,  $\mathbf{e}_{2}=-\frac{1}{2}\hat{\mathbf {x}} + \frac{\sqrt{3}}{2}\hat{\mathbf{y}}$ and $\mathbf{e}_{3}=-\frac{1}{2} \hat{\mathbf{x}} - \frac{\sqrt{3}}{2}\hat{\mathbf{y}}$. For $\frac{J_1}{J_2} > 1/3$, the minimal-energy state in zero magnetic field is a spiral with spins rotating in a vertical plane, which transforms into multiply-periodic states in applied magnetic fields.\cite{Leonov2015} Strong magnetic fields and large easy-axis anisotropy supress all modulated states, but they can still allow for metastable isolated skyrmions, which is the region of parameters considered here.

\begin{figure}
	\centering	
	\includegraphics[width=0.9\linewidth]{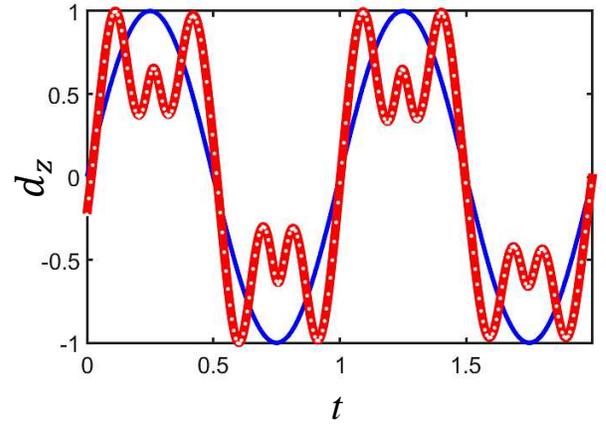}
	\caption{(Color online) Time dependence of the $z$-component of the skyrmion electric dipole moment $d_z(t) = d_z(t)$ (thick line) in the oscillating electric field $E_z(t) = E_0 \cos (\omega t + \varphi_0)$ (thin line). The time $t$ is measured in units of $T = \frac{2\pi}{\omega}$ -- the oscillation period of the electric field; $d_{z}(t)$ and $E_{z}(t)$ are divided by their maximal values. White dots show the fit of $d_z(t)$ by $\left(d_0 + d_1\mu_z(t)\right) \cos \chi(t)$, where $\chi(t)$ is the skyrmion helicity angle, $\mu_z(t)$ is the $z$-component of its magnetic moment, and $d_{0}$ and $d_{1}$ are the fitting parameters.}
	\label{fig:deltadzEz}
\end{figure}

\begin{figure}
	\centering
	\includegraphics[width=0.9\linewidth]{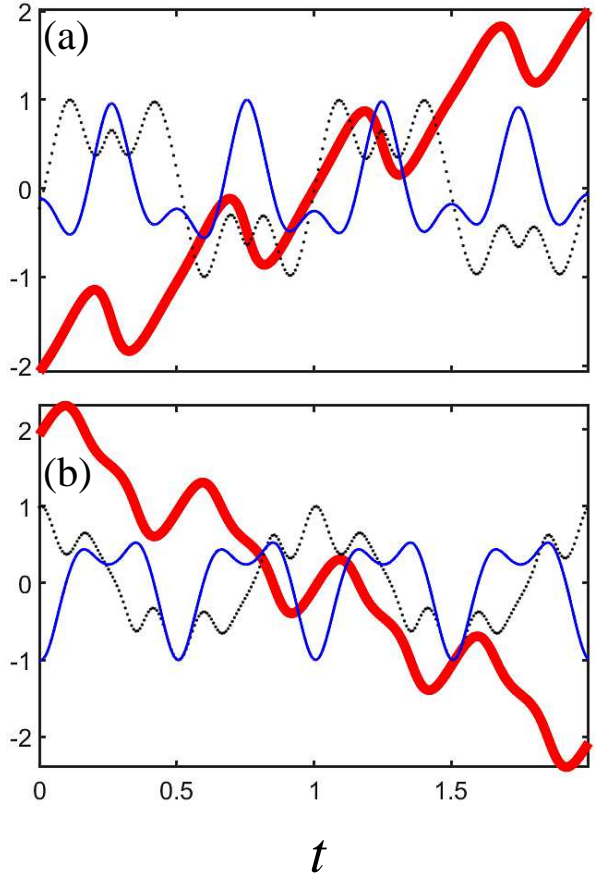}
	\caption{(Color online) (a) Time dependence of the skyrmion helicity angle $\chi(t)$ measured in units of $\pi$ (thick solid line) in the periodically oscillating electric field, for the anticlockwise rotation of the in-plane spin components. During half-period of the electric field oscillations, $T/2$, $\chi$ increases by $\pi$. Also shown are the $z$-components of the skyrmion magnetic moment $\delta \mu_z(t) = \mu_z(t) - \langle \mu_z \rangle$ (thin solid line) and electric dipole moment $d_z(t)$ (dotted line). (b) The same for clockwise spin rotation.}
	\label{fig:chiupchidown}
\end{figure}

Figures~\ref{fig:deltadzEz} and \ref{fig:chiupchidown} show the results of numerical solution of LLG equation describing spin dynamics for the model parameters $J_1 = 1, J_2 = 0.36, h_z = 0.05, K = 0.1$, $E_0 = 0.009$, $\omega = 0.012$, and the Gilbert damping parameter $\alpha = 0.003$. Figure~\ref{fig:deltadzEz}  shows time dependence of the skyrmion electric dipole moment $d_z(t)$ and the electric field $E_z(t)=E_0 \cos (\omega t + \varphi_E)$ (both normalized by their maximal values). The spin rotation period coincides with the oscillation period of $E_z(t)$, $T = \frac{2\pi}{\omega}$ and spins rotate in such a way that $d_z(t)$ oscillates in phase with $E_z(t)$. 

The skyrmion helicity is defined by $\chi = \arctan \left(\frac{A_y}{A_x}\right)$, where
\begin{equation}
\left\{
\begin{array}{ccl}
A_x & = & \sum\limits_{n,\alpha} \left[\mathbf{e}_\alpha \times \hat{\mathbf{ z}}\right]\cdot \left[\mathbf{m}_n \times 
\mathbf{m}_{n+e_\alpha}\right], \\ \\
A_y & = & \sum\limits_{n,\alpha} \mathbf{e}_\alpha \cdot \left[\mathbf{m}_n \times \mathbf{m}_{n+e_\alpha}\right].
\end{array}
\right.
\end{equation}
This definition can be justified as follows. In the continuum limit, $A_x \propto - \int\!\! d^2x  \, m_j \overleftrightarrow{\partial_j} m_z$ and  $A_y \propto - \int\!\! d^2x \left(\mathbf{m} \cdot \mathbf{\nabla} \times \mathbf{m}\right)$. For the skyrmion configuration Eq.(\ref{eq:skyrmion}) with $\rm v = +1$, $(A_x,A_y) = \frac{\sqrt{3}D}{\lambda_\perp}(\cos\chi,\sin\chi)$, where $D$ is defined by Eq.(\ref{eq:dz}).

After initial back-and-forth oscillations, spins start rotating clockwise or anticlockwise, depending on the initial skyrmion helicity and phase $\varphi_E$ of the electric field oscillations.  Figures~\ref{fig:chiupchidown}(a) and (b) show the time dependence of the skyrmion helicity (thick solid line) in the steady state with the anticlockwise and clockwise spin rotation directions, respectively. The helicity angle $\chi$ is given in units of $\pi$. As the time $t$ increases by $T/2$, both $E_z$ and $d_z$ change sign, while $\chi$ increases or decreases by $\pi$. 

The time dependence of $\chi$ is non-monotonic: the rotation direction temporarily reverses near $\chi = 0$ and $\chi = \pi$, which gives rise to oscillations of the skyrmion radius and its magnetic moment $\mu_z = \sum_i (m_{iz}- 1)$ counted from the magnetic moment of the ferromagnetic state with $m_z = +1$. For $h_z > 0$, $m_z$ near the skyrmion center is negative and a larger $\mu_z$ corresponds to a smaller skyrmion radius. 

The skyrmion helicity angle $\chi$ and the magnetic moment $\mu_z$ are canonically conjugated variables and the coupling between their dynamics provides an effective mass term in the equation for $\chi$ \cite{Leonov2017}. This inertia is necessary for the persistent rotation in one direction. Without it, $\chi$ would oscillate together with the electric field (which it does, if the amplitude $E_0$ of the electric field is too small). The $\mu_z$-oscillations affect the magnitude of the electric dipole moment $D$ in Eq.(\ref{eq:dz}) that to a high precision is a linear function of $\mu_z$: $D = d_0 + d_1 \mu_z$ (white dots in Fig.~\ref{fig:deltadzEz} show $(d_0 + d_1\mu_z(t))\cos\chi(t)$). The time evolutions of $\chi$ (thick solid line), $\mu_z$ (thin solid line) and $d_z$ (dotted line) for clockwise and anticlockwise rotations shown in Figs.~\ref{fig:chiupchidown}(a) and (b), though similar, are not symmetry related. In fact, for a fixed magnetic field $h_z$, no symmetry of the LLG equation can transform the steady-state rotation in the clockwise direction into the anticlockwise rotation.  

For small Gilbert damping parameter $\alpha$, the time dependence of $\chi$, $\mu_z$ and $d_z$ in the steady state becomes independent of $\alpha$, which is different from the dynamics of a ferromagnetic domain wall in an applied magnetic field that moves with a velocity inversely proportional to $\alpha$, as  required by energy conservation.\cite{Walker1974} At larger $\alpha$, the skyrmion electric dipole moment $d_z$ does not oscillate in phase with the electric field and the energy supplied at the  rate $- d_z \frac{dE_z}{dt}$ is partly carried away from the skyrmion by radially propagating spin waves emitted by the rotating spins.

\begin{figure}
	\centering
	\includegraphics[width=0.9\linewidth]{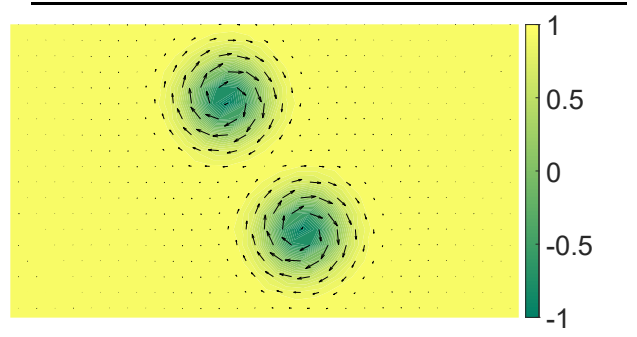}
	\caption{(Color online) A pair of skyrmions on a triangular spin lattice with nearest-neighbor ferromagnetic and next-nearest-neighbor antiferromagnetic Heisenberg exchange interactions (see text). In-plane spin components are shown by arrows; the out-of-plane spin component is shown by contour plot.}
	\label{fig:Skpair}
\end{figure}

\section{Rotation of skyrmions in oscillating electric field}
\label{sec:skskpair}

\begin{figure}
	\centering
	\includegraphics[width=0.9\linewidth]{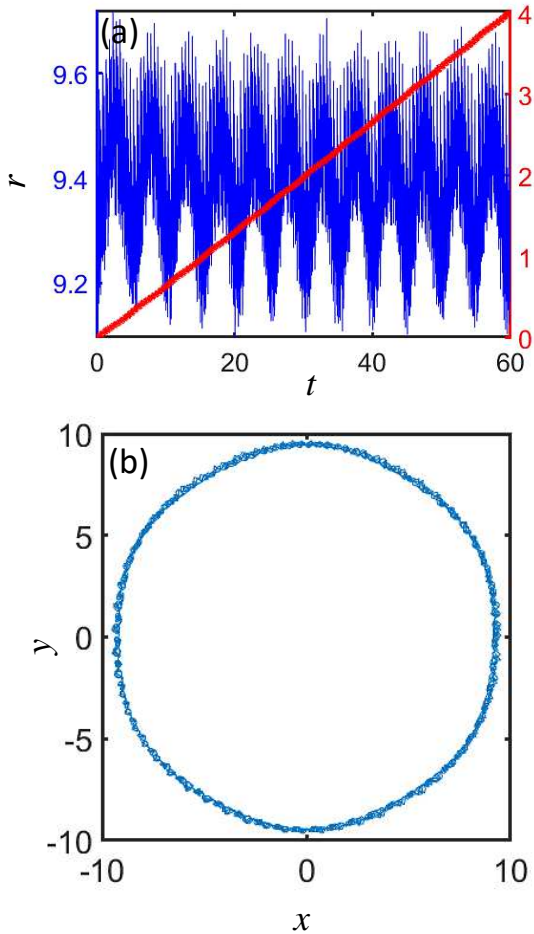}
	\caption{(Color online) (a) Time evolution of the distance $r$ between two skyrmions  in the periodically oscillating electric field (thin line; the values in units of the lattice constant are given on the left side of the plot) and of the angle $\psi$  (thick line; the values in units of $\pi$ are given on the right side of the plot) describing their rotation around each other. The time $t$ is measured in units of the electric field oscillation period $T$. The period of rotation of skyrmions around each other is $30 T$. (b) The rotation trajectory: $y = r \sin \psi$ vs $x = r \cos \psi$.}
	\label{fig:Skskrotation}
\end{figure}

In frustrated magnets, the angle $\theta$ describing skyrmion spin texture Eq.(\ref{eq:skyrmion}) is a nonmonotonic function of the distance from the skyrmion center. \cite{Leonov2017} These oscillations give rise to oscillations of the interaction energy $U$ of two skyrmions as a function of the distance $r$ between them. In addition, $U$ depends on helicities of the skyrmions.\cite{Leonov2017} If two skyrmions with the same helicity are placed in a local minimum of $U(r)$ (see Fig.~\ref{fig:Skpair}) and are subjected to a periodically oscillating $E_z(t)$, the in-plane spin components of both skyrmions start rotating and the coupling between helicity and translational modes leads to rotation of the skyrmions around each other.

Figure~\ref{fig:Skskrotation}(a) shows time dependence of the distance $r$ between the skyrmions (thin line) and the angle $\psi$ describing the skyrmion rotation (thick line) calculated for the model parameters $J_1,J_2,h_z,K$, and $\omega$ used in the previous section, $E_0 = 0.04$ and $\alpha = 0.02$. The skyrmion coordinates, $\mathbf{R}_i = (X_i,Y_i), i=1,2$,  are the interpolated positions of the points where $m_z = -1$, $\mathbf{r} = (x,y) = \mathbf{R}_1 - \mathbf{R}_2$ is the relative coordinate vector and $\psi = \arctan \left(\frac{y}{x}\right)$. The thickness of $r$-plot originates from multiple spin rotations, which result in oscillations of the distance between skyrmions: the skyrmion rotation period equals 30 spin-precession periods. One can also see 6 slow modulations of $r$ during one skyrmion rotation and the skyrmions rotate faster when the distance between them is smaller. These distance variations result from the six-fold modulation of the ideal rotationally symmetric skyrmion spin configuration caused by the triangular lattice. A hexagonal distortion of a circular skyrmion trajectory can also be seen in Fig.~\ref{fig:Skskrotation}(b), where $y(t)$ is plotted versus $x(t)$.

We note that skyrmions also rotate when the distance $r$ between them monotonically increases or decreases towards an optimal distance, at which their interaction energy is minimal. However, such a rotation stops when the energy minimum is reached. By contrast, the skyrmion rotation driven by the oscillating electric field does not change the average $r$, as can be seen from Fig.~\ref{fig:Skskrotation}, and it does not stop. This rotational motion is somewhat chaotic due to the emission of spin waves and lattice pinning, which `broadens' the trajectory line shown in Fig.~\ref{fig:Skskrotation}(b).  

\section{Translational motion of the skyrmion-antiskyrmion pair in oscillating electric field}
\label{sec:skaskpair}

\begin{figure}
	\centering
	\includegraphics[width=0.9\linewidth]{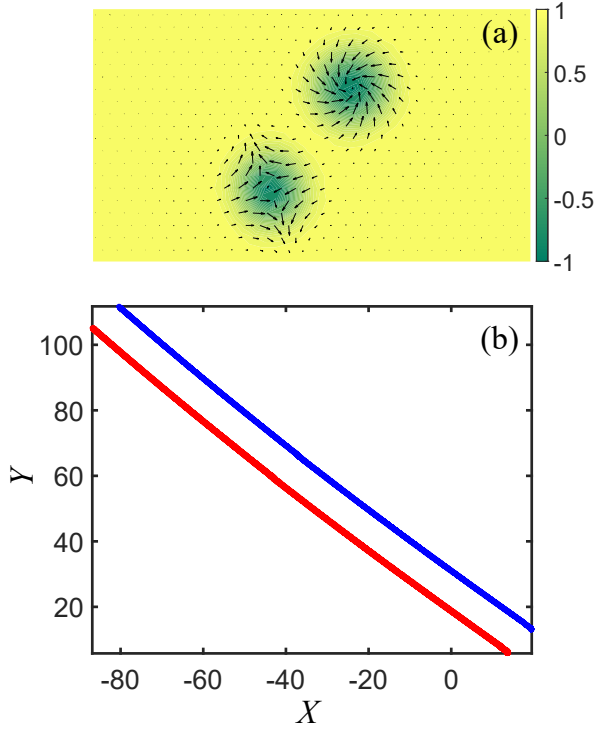}
	\caption{(Color online) (a) The skyrmion-antiskyrmion pair in a frustrated triangular antiferromagnet. In-plane spin components are shown with arrows; the out-of-plane spin component  is shown by contour plot. (b) Skyrmion (upper line) and antiskyrmion (lower line) trajectories traversed during $\sim 270$ oscillations of the electric field applied in the $z$-direction. The coordinates of the topological defects are measured in units of the lattice constant.}
	\label{fig:Skaskpath}
\end{figure}

Most interestingly, if one of the skyrmions is replaced by an antiskyrmion (see Fig.~\ref{fig:Skaskpath}(a)), the coupling of spins to an oscillating electric field results in a translational motion of the  skyrmion-antiskyrmion pair. As discussed in Sect.~\ref{sec:spinrotation}, the electric dipole moment of antiskyrmion is zero and spins in an isolated antiskyrmion do not rotate in an applied electric field. However, when it forms a pair with a skyrmion, spins of both topological defects rotate with the same frequency $\omega$ equal to the frequency of oscillating electric field. While a pair of skyrmions rotates around a fixed point, the skyrmion-antiskyrmion pair rotates very little. Instead, it moves in the direction approximately normal to the relative coordinate vector $\mathbf{r}$. Figure~\ref{fig:Skaskpath}(b) shows the trajectories of skyrmion (upper line) and antiskyrmion (lower line) traversed over the time equal to  $\sim 230$ spin rotation periods, during which the pair moves by $\sim 140$ lattice constants. The average distance $r$ between the topological defects does not change with time and, hence, this translational motion cannot result from a decreasing potential energy. This calculation was performed for the spin-model parameters used in the previous two sections, the amplitude $E_0 = 0.04$ and frequency $\omega = 0.012$ of the electric field, and the damping parameter $\alpha = 0.008$. 

The skyrmion and antiskyrmion trajectories are not perfectly straight due to a slow rotation of the pair: the relative coordinate vector $\mathbf{r}$ rotates through an angle of about $- 6^\circ$ over the whole time of motion and so does the center-of-mass velocity of the pair,  $\frac{1}{2}\left(\dot{\mathbf{X}}_1 + \dot{\mathbf{X}}_2\right)$, i.e. the pair moves along a circle of large radius.

\section{Discussion and Summary}
\label{sec:conclusions}

The electric field-induced dynamics of the skyrmion-skyrmion and skyrmion-antiskyrmion pairs described above are complex. Strong sensitivity to the electric field frequency found in numerical simulations indicates   excitation of modes corresponding to oscillations of the relative distance and helicity of the two topological defects. However, the marked difference between the rotation of the skyrmion pair and translational motion of the skyrmion-antiskyrmion pair may have topological origin, namely the skyrmion Hall effect, i.e. the skyrmion motion in a direction transverse to the electron \cite{Jiang2017} or magnon \cite{Nagaosa2013,Mochizuki2014} current. 


Consider a pair of skyrmions with topological charges $Q_1 = Q_2 = Q$ and coordinates $\mathbf{R}_i = (X_i,Y_i)$, $i = 1,2$, and assume that initially $Y_1 = Y_2 = 0$. The dynamics of skyrmion coordinates is described by Thiele equations\cite{Mochizuki2014,Iwasaki2013}
\begin{equation}
		\left\{
	\begin{array}{cl}
	\alpha \Gamma \dot{X}_i - G \dot{Y}_i &\!\!\!\!\!= \pm F_x,  \\
	G\dot{X}_i + \alpha \Gamma \dot{Y}_i &\!\!\!\!\!= \pm F_y, 
	\end{array}
	\right.
	\label{eq:sseqmotion}
\end{equation}
where $G = 4 \pi Q$, $\Gamma = \int \!\!d^2x (\partial_x \mathbf{m})^2 = 
\int \!\!d^2x (\partial_y \mathbf{m})^2$ and the sign in the right-hand side of the equations is $+$ for $i = 1$ and $-$ for $i = 2$. 

Although it seems obvious that the force acting on skyrmion 1 should be equal to minus the force acting on skyrmion 2,
$\mathbf{F}_1 = (F_x,F_y) = - \mathbf{F}_2$, it is worthwhile to discuss this in detail, since the forces are partly non-potential. The force $F_y$, perpendicular to the relative coordinate vector $\mathbf{r} = \mathbf{R}_1 - \mathbf{R}_2$, results from the topological magnon Hall effect.\cite{Nagaosa2013,Mochizuki2014} Magnons, emitted by one skyrmion due to the electrically-excited spin rotation, skew-scatter off the effective magnetic field of another skyrmion and acquire a transverse momentum, which exerts a reaction force on the latter skyrmion. The total flux of the effective magnetic field, $\Phi = Q \Phi_0$, $\Phi_0$ being the magnetic flux quantum, is proportional to the skyrmion topological charge and so is the reaction force $F_y$: $F_y = - A I Q$. Here, $I = I_{2 \rightarrow 1}$ is the magnon current that skyrmion 2 shines on skyrmion 1 and $A$ is a coefficient. The magnon scattering also results in a transfer of the $x$-component of the magnon momentum to skyrmion 1, which is independent of $Q$: $F_x = B I + f$, the second term being the central force due to the interaction $U(X_1-X_2)$ between the skyrmions: $f = - \frac{\partial U}{\partial X_1}$. The force acting on skyrmion 2 has an opposite direction, because of the change of magnon current and central force directions, $I_{1 \rightarrow 2} = - I_{2 \rightarrow 1}$, $f \rightarrow -f$, 
\textit{and} because of equal topological charges of the two skyrmions.

In this case, the `center-of-mass' coordinates of the skyrmion pair, $\mathbf{R} = \frac{1}{2}\left(\mathbf{R}_1 + \mathbf{R}_2\right)$, are constants of motion and the relative coordinate dynamics is described by
\begin{equation}
	\left\{
	\begin{array}{cl}
		\alpha \Gamma \dot{r} - G r \dot{\psi}    &\!\!\!\!\!= 2 F_r = 2(B I + f), \\
		G\dot{r}_i + \alpha \Gamma r \dot{\psi}_i &\!\!\!\!\!= 2 F_y = - 2 A I Q,
	\end{array}
	\right.
	\label{eq:rdynamics}
\end{equation}
where the Cartesian coordinates $(x,y)$ are replaced by polar coordinates $(r,\psi)$.

The forces in radial and azimuthal directions depend on the distance $r$ between skyrmions, which remains constant, if 
\begin{equation}
	\frac{F_r}{F_\psi} = - \frac{G}{\alpha\Gamma} = -\frac{8\pi A I}{\alpha \Gamma},
	\label{eq:forceratio}
\end{equation}
where we used $Q^2  = 1$. It may seem that Eq.(\ref{eq:forceratio}) cannot hold for small damping constants. However, as $\alpha$ decreases, so does the emitted magnon current $I$, since for $\alpha \ll 1$, the skyrmion electric dipole moment oscillates in phase with the electric field (see Fig.~\ref{fig:deltadzEz}). 

For $\dot{r} = 0$, the skyrmion rotation rate is given by
\begin{equation}
	\dot{\psi} =- Q \frac{2 A I}{\alpha \Gamma r}.
\end{equation}
Since  $A>0$, skyrmions rotate clockwise(anti-clockwise), for $Q = +1(Q = - 1)$. The rotation of the skyrmion pair is similar to the rotation of skyrmion crystal induced by the thermal magnon current,\cite{Mochizuki2014} except that in our case skyrmions themselves are the sources of magnons. 

We now replace skyrmion 2 by antiskyrmion with $Q_2 = - Q$. Thiele equations for the antiskyrmion,
\begin{equation}
	\left\{
	\begin{array}{cl}
		\alpha \Gamma \dot{X}_i + G \dot{Y}_i &\!\!\!\!\!= - F_x, \\
		- G\dot{X}_i + \alpha \Gamma \dot{Y}_i &\!\!\!\!\!= + F_y,
	\end{array}
	\right.
\end{equation}
are obtained from those for skyrmion 2 (Eqs.(\ref{eq:sseqmotion}) with the minus sign in the right-hand side) by changing the sign of $G$ and $F_y$ that are proportional to topological charge. Importantly, the transverse force $F_y$ acting on antiskyrmion equals that for skyrmion, since both the direction of magnon current and topological charge changed sign. We assume that $I_{1 \rightarrow 2} = - I_{2 \rightarrow 1}$, since spins in the skyrmion  and antiskyrmion rotate at the same rate. 

The center-of-mass and relative dynamics are now coupled:
\begin{equation}
	\left\{
	\begin{array}{cl}
		\alpha \Gamma \dot{X}_i + G \dot{y}_i  &\!\!\!\!\!= 0, \\
		- G\dot{X}_i + \alpha \Gamma \dot{y}_i &\!\!\!\!\!= 0, 
	\end{array}
	\right.
\end{equation}
and 
\begin{equation}
	\left\{
	\begin{array}{cl}
		\alpha \Gamma \dot{x}_i + G \dot{Y}_i  &\!\!\!\!\!= 2 F_x, \\
		- G\dot{x}_i + \alpha \Gamma \dot{Y}_i &\!\!\!\!\!= 2 F_y. 
	\end{array}
	\right.
\end{equation}
$(X,y)$ and $(x,Y)$ are two pairs of canonically conjugated dynamical variables, which also follows from the fact that the kinetic part of the Lagrangian of skyrmion-antiskyrmion pair, $G X_1 \dot{Y}_1 - G X_2 \dot{Y}_2$, equals
$G X \dot{y}_1 + G x \dot{Y}_2$.

The constants of motions are $X$ (the pair can only move in the direction perpendicular to the skyrmion-antiskyrmion `bond') and $y$ (the pair does not rotate). If the distance $x$ between  skyrmion and antiskyrmion is constant, the pair moves in the $y$-direction with the velocity $\dot{Y} = - Q \frac{2 A I}{\alpha \Gamma}$, where $Q$ is the topological charge of skyrmion. 

This behavior is consistent with the results of numerical simulations. The small deviation of the angle between the center-of-mass velocity of the pair $\dot{\mathbf{R}}$ and  relative coordinate vector $\mathbf{r}$ from $90^\circ$, and the slow rotation of $\mathbf{r}$ found in numerical simulations may be related to a difference between the magnon currents, $|I_{1 \rightarrow 2}|$ and $|I_{2 \rightarrow 1}|$. Furthermore, we neglected oscillations of the distance $r$ between the topological defects, as well as variations of their size and shape. 

The number of skyrmion materials is rapidly growing. \cite{Kanazawa2021} Yet, magnetically frustrated Mott insulators that can host skyrmions and show the effects discussed in this paper are still to be found. The magnetoelectric coupling required to excite the dynamics of topological defects has to be strong enough to overcome the helicity pinning by an in-plane magnetic anisotropy and magnetodipolar interactions. The in-plane anisotropy is weak in hexagonal materials, such as transition metal halides,\cite{McGuire2017} where it is propotional to sixth power of the spin-orbit coupling. Long-ranged interactions between magnetic dipoles favor Bloch skyrmions.  This magnetic anisotropy is comparable with the energy of  skyrmion electric dipole in the electric field $E_z = 10^5$ V$\cdot$cm$^{-1}$, for the magnetization $M = 100$ G and the magnetoelectric coupling strength of TbMnO$_3$.\cite{Kimura2003} The anisotropy resulting from the magnetodipolar interactions is reduced in antiferromagnets, whereas the magnetoelectric energy is not. The N\'eel vector $\mathbf{l}$ describing a collinear antiferromagnetic ordering transforms as $\mathbf{m}$, except for an additional minus sign when spin-up and spin-down magnetic sublattices are interchanged by a crystal symmetry  transformation. However, this sign is canceled in the magnetoelectric coupling Eq.(\ref{eq:fmexy}), where $\mathbf{m}$ is replaced by $\mathbf{l}$. Therefore, symmetry requirements for the electric polarization induced by ferromagnetic and antiferromagnetic spin textures are the same. 

To summarize, the electric dipole moments and charges induced by skyrmions and merons allow for the control of topological magnetic defects in multiferroic materials with an applied electric field. Oppositely charged vortices and antivortices shift in an electric field in opposite directions. Skyrmions are electrically neutral, but they posses an out-of-plane electric dipole moment that depends on skyrmion helicity, which leads to spin procession in an oscillating electric field. The coupled helicity and translational dynamics of skyrmions make possible to rotate skyrmions forming pairs and move skyrmion-antiskyrmion pairs.

The voltage control of topological defects in magnetic insulators discussed in this paper can be interesting for low energy-consumption memory and data processing devices. Skyrmions with spins rotating in an oscillating electric field  can act as spin-wave generators playing a role of `active bits' that can force neighboring bits to move. The skyrmion pair rotation can be used for electrical reshuffling of skyrmion positions, e.g., in logical operations. Numerical simulations show that an oscillating electric field can also rotate small clusters of skyrmions. The translational dynamics of skyrmion-antiskyrmion pairs can be employed to shift topological defects over long distances.

\begin{acknowledgment}
	
	The author is grateful to N. Nagaosa and Y. Tokura for inspiring discussions. This work was supported by Vrije FOM-programma ‘Skyrmionics’.
	
\end{acknowledgment}

\end{document}